\documentclass[11pt,fleqn,twosides]{article}

\usepackage{latexsym}
\usepackage[tbtags]{amsmath}
\usepackage{amsopn,amsthm,amsfonts,amssymb}
\usepackage{fancyhdr}
\usepackage{latexsym}
\fancypagestyle{plain}{ \fancyhead{} \fancyfoot{}

}

\newcommand{\mc}[1]{\ensuremath{\mathcal{#1}}}

\newcommand{\beq}{\begin{equation}}
\newcommand{\eeq}{\end{equation}}
\newcommand{\beqa}{\begin{eqnarray}}
\newcommand{\eeqa}{\end{eqnarray}}
\newcommand{\sm}[1]{\ensuremath{m_{(#1)}}}

\newtheorem{theorem}{Theorem}
\newtheorem{lemma}[theorem]{Lemma}
\newtheorem{proposition}[theorem]{Proposition}

\newtheorem{definition}[theorem]{Definition}

\theoremstyle{remark}

\newtheorem*{acknow}{Acknowledgments}
\let\d\partial
\let\n\noindent
\let\la\lambda
\def\ds{\displaystyle}
\def\D{{\mathcal D}}

\let\la\lambda
\let\La\Lambda

\let\om\omega
\let\da\dagger
\let\Om\Omega
\let\n\noindent
\let\d\partial
\let\y\infty
\let\rw\rightarrow

\begin{document}
\small

\title{Generalized Hermite polynomials in superspace as eigenfunctions
of the supersymmetric rational CMS model}

\author{Patrick Desrosiers\thanks{pdesrosi@phy.ulaval.ca} \cr
\emph{D\'epartement de Physique, de G\'enie Physique et
d'Optique},\cr Universit\'e Laval, \cr Qu\'ebec, Canada, G1K 7P4.
\and Luc Lapointe\thanks{lapointe@inst-mat.utalca.cl }\cr
\emph{Instituto de Matem\'atica y F\'{\i}sica},\cr Universidad de
Talca,\cr Casilla 747, Talca, Chile.
  \and
  Pierre
Mathieu\thanks{pmathieu@phy.ulaval.ca} \cr \emph{D\'epartement de
Physique, de G\'enie Physique et d'Optique},\cr Universit\'e
Laval, \cr Qu\'ebec, Canada, G1K 7P4. }

\date{May 2003}

\maketitle

\begin{abstract}

We present an algebraic construction of the orthogonal
eigenfunctions of the supersymmetric extension of the rational
Calogero-Moser-Sutherland model  with harmonic confinement.  These
eigenfunctions are the superspace extension of the generalized
Hermite (or Hi-Jack) polynomials.  The conserved quantities of the
rational supersymmetric  model are  related to their
trigonometric relatives through a similarity transformation. This
leads to a simple expression between the corresponding
eigenfunctions: the generalized Hermite superpolynomials are written as a
differential operator acting  on the corresponding Jack superpolynomials.
As an aside, the maximal superintegrability of the supersymmetric rational
Calogero-Moser-Sutherland model is demonstrated.

\end{abstract}

\newpage

 \tableofcontents

\vfill\break

\section{Introduction}

Operator constructions of the eigenfunctions of the rational
Calogero-Moser-Sutherland\footnote{We use the qualitative
`Calogero-Moser-Sutherland' to describe the generic class of
models that includes the models studied by Calogero and Sutherland
in the quantum case  and by Moser in the classical case. In this
work, we only treat the quantum case,  and in this context the
name of Moser is often omitted. The rationale for its inclusion is
due to the fundamental importance of the Lax formulation in the
quantum case, which is a direct extension of the classical one
that he introduced. Since the quantum rational model with
confinement was studied by Calogero, what we call the rCMS model
is often referred to as the Calogero model.} model with
confinement (rCMS) \cite{CMS},
   \begin{equation}
\mc{H}= \frac12\sum_{i=1}^N \left[-\frac{\d^2}{ \d
x_i^2}+\om^2x_i^2\right] + \sum_{1\leq i<j\leq
N}\frac{\beta(\beta-1)}{ x_{ij}^2},\end{equation} were first
initiated by Perelomov \cite{Pere}, who tried to obtain  creation
operators $B_n^\da$ satisfying $[ \mc{H},B_n^\da]=n\om B_n^\da $
in order to obtain the excited states as $|\la\rangle=
\prod_{i=1}^N (B_n^\da)^{\la_i}|0\rangle$. However, his technique
only allowed him to obtain explicitly the first few $B_n^\da$. A
complete set of creation operators was later presented in
\cite{Brink}, using the powerful (and at the time, rather new)
Dunkl-operator formalism. An independent construction (later found to be
related to the former in a simple way
\cite{UWa2}) was also presented in
\cite{UWa1}, in which  the creation operators were expressed directly in
terms of the Lax operators.  However, the
resulting eigenfunctions are not orthogonal.

The first few orthogonal eigenfunctions were constructed in
\cite{UWa3} by the simultaneous diagonalization of the Hamiltonian
and the first non-trivial conserved operator. In this way, no
pattern ever showed up. Shortly after, a quite different approach
to the construction of the rCMS eigenfunctions was presented in
\cite{UWa4}.  It consisted in the extension to the rational case
of the operator method  found in \cite{Lap} for building the
excited states (the Jack polynomials) of the trigonometric CMS
(tCMS) model. That this operator method can be translated directly
to the rational case is certainly remarkable and actually rather
surprising at first sight.\footnote{There is of course an extra
parameter in the rational case, namely $\om$.   In the absence of
confinement $(\om=0)$, the rational case can be recovered as a
limiting case of the trigonometric model.  The above situation is
thus a priori unexpected.}  The first few orthogonal functions
constructed in \cite{UWa3} were  recovered using this operator
method. In the $\om\rw \y$ limit, the rCMS eigenfunction indexed
by the partition $\la$ and written $J_\la^\om$ reduces to the
corresponding Jack polynomial $J_\la$:
\begin{equation}
  \lim_{\omega\rightarrow\infty}J_\la^\om=J_\la.
\end{equation}
  Because of
this {\it hi}dden connection with the Jack polynomials,  the
orthogonal rCMS eigenfunctions have been called the {\it Hi-Jack
polynomials}.\footnote{More recently, a second orthogonal basis
has been found in \cite{UWa5}, building on the work \cite{GP}.}

In parallel to these developments in physics, the very same
functions were studied in mathematics under the name of {\it
generalized Hermite polynomials}. Indeed, when $N=1$, the
eigenfunctions of the rCMS Hamiltonian are precisely the Hermite
polynomials. For $\beta=1/2$, the generalized Hermite polynomials
have first been introduced in \cite{James}. The general case was
treated by Lasalle in \cite{Las}. There, the generalized Hermite
polynomials are defined uniquely
   (up to normalization) from
   their eigenfunction characterization (in terms of a differential
operator equivalent to the
rCMS Hamiltonian without the ground-state contribution) and their
   triangular decomposition in terms  of
    Jack polynomials.  Lasalle also
found a remarkably simple  operational relation between the Jack
   and the generalized Hermite/Hi-Jack polynomials:\footnote{A later
derivation of this formula was given by Sogo
     in \cite{Sog}.}
\begin{equation}\label{formHermite}
  J_\la^\om= e^{-\Delta/4\om} J_\la,
\end{equation}
where $\Delta$ is some operator to be defined below. This provides
a direct one-to-one correspondence between Jack polynomials and
their `{\it hi}gher' relatives and also, in principle, a
computational tool for constructing the $J_\la^\om$'s.  A
rather extensive analysis of their properties is presented in
\cite{vandiejen, BF}.\footnote{Reference \cite{BF} also mentions
two unpublished manuscripts, one by Macdonald and the other by
Lasalle, where further properties of these generalized Hermite
polynomials are presented.}

In this work, we  present a very natural extension of these
polynomials, by constructing  their superspace analogues.  The
resulting {\it generalized Hermite  superpolynomials} (also called
the generalized Hermite polynomials in superspace) reduce to the
usual generalized Hermite polynomials in the zero-fermion sector
(the fermion-sector label ranging
  from $0$ to $N$).  The generalized Hermite
superpolynomials  are eigenfunctions of
the supersymmetric rCMS (srCMS) model first introduced in
\cite{Freedman}.\footnote{An algebraic construction of the
eigenstates that generalizes  Perelomov's  was proposed in
\cite{Freedman}, but only the first non-trivial creation operators
was explicitly written.  Later, Ghosh \cite{ghosh} obtained all
fermionic and bosonic operators using a similarity transformation
from the srCMS model to  supersymmetric harmonic oscillators. His
set of solutions was however over-complete. Note that
superpartitions, which provide the proper solution labelling, were
note known at the time. } Our main result is to provide
the direct supersymmetric lift of the Lasalle's operational
definition.  The generalized Hermite
superpolynomials are then defined in terms of the
  the Jack superpolynomials, which are themseves
  eigenfunctions of the supersymmetric extension of the
tCMS model \cite{DLM1,DLM2,DLM3}.\footnote{The
relationship between the three articles \cite{DLM1, DLM2, DLM3} is
the following. In \cite{DLM1}, we construct eigenfunctions of the
stCMS Hamiltonian that decompose triangularly in the supermonomial
basis. These eigenfunctions were called Jack superpolynomials (and
denoted ${\cal J}_\La$) because they reduce to Jack polynomials in
the zero-fermion sector. In \cite{DLM2}, the coefficients in their
triangular decomposition have been computed explicitly, leading to
a determinantal expression for the ${\cal J}_\La$'s. The ${\cal
J}_\La$'s diagonalize the $N$ bosonic conserved charges that
reduce to the tCMS conserved quantities in absence of fermionic
variables \cite{DLM3}. However they are not orthogonal. To
construct orthogonal combinations of the ${\cal J}_\La$'s, we
needed to find the eigenfunctions of the second tower of bosonic
conserved charges. This was done in \cite{DLM3}. The resulting
orthogonal eigenfunctions are written $J_\La$. These are the
superpolynomials that, from our point of view,
  deserve to be called the Jack
superpolynomials. From now on, we thus use the name `Jack
superpolynomials' or equivalently,  `Jack polynomials in
superspace', for the $J_\La$'s of \cite{DLM3}.  A survey of these
three articles is presented in \cite{DLM4}.}

As pointed out in the conclusion, there are other ways of
constructing the generalized Hermite superpolynomials. However, in
this work, we  focus  on the construction that is
closely tied to the underlying physical Hamiltonian.

The article is organized as follows. Some relevant background
material in presented in section 2. In particular, we give the
definition of the srCMS Hamiltonian  as well as that of the
symmetric superfunctions. We then display three bases for the ring
of superfunctions, the monomial symmetric superfunctions (called
the supermonomials),  the Jack superpolynomials and the power-sum
superpolynomials.

Sections 3 and 4 is concerned with the superextension of Lasalle's
construction. This  is embedded within a broader analysis of the
srCMS algebraic structure centered on the Dunkl operator and the
relationship between the srCMS and stCMS conserved quantities.  In
particular, it is shown that all srCMS conserved quantities can be
readily obtained as `gauge transformations' of the stCMS ones.  As
an offshoot of these algebraic considerations, we establish the
superintegrability of the srCMS model in Section 5 and display two
other bases
  of eigenfunctions in Section 6.

A summary of our main results, concluding remarks as well as a
listing of some natural extensions are collected in Section 7.

\section{The srCMS model and symmetric functions in superspace}

In this section, we first define the srCMS model, and then
introduce the essential  quantities, such as superpartitions and
bases of superpolynomials, that will be needed to construct its
eigenfunctions.

\subsection{The  srCMS Hamiltonian}

The Hamiltonian of the srCMS model reads \cite{Freedman}:
     \begin{equation}
\mc{H}= \frac12\sum_{i=1}^N \left[-\frac{\d^2}{ \d
x_i^2}+\om^2x_i^2\right] + \sum_{1\leq i<j\leq
N}\frac{\beta(\beta-1+\theta_{ij}\theta_{ij}^\dagger)}{
x_{ij}^2}-\frac{N}{2}[\beta(N-1)-\omega]\end{equation} where \beq
x_{ij}=x_i-x_j,\qquad\theta_{ij} = \theta_{i} -
\theta_{j}\qquad\mbox{and}\qquad \theta_{ij}^{\dagger} =
\partial_{\theta_{i}} - \partial_{\theta_{j}}.\eeq
  The
$\theta_{i}$'s, with  $ i=1,\cdots,N $, are anticommuting
(Grassmannian or fermionic) variables.  The commuting (bosonic)
variables $x_j$ are real. The parameters $\beta$ and $\om$  also
belong to the real field.

The model is supersymmetric since it can be written as the
anticommutation of two
  charges,   \beq
\mc{H}=\frac{1}{2}\{Q,Q^\dagger\}\qquad\mbox{with}\qquad
Q^2=(Q^\dagger)^2=0,\eeq where the srCMS fermionic operators  are
\beqa\label{fcharges}
Q\phantom{^\dagger}&=&\sum_{j=1}^N\theta_j\,\,\left(\phantom{-}\partial_{x_j}+\om
x_j-\beta\sum_{1\leq k \leq N \atop k\neq
j}\frac{1}{x_{jk}}\right),\cr Q^\dagger&=&\sum_{j=1}^N
\partial_{\theta_j}\left(-\partial_{x_j}+\om x_j-\beta\sum_{1\leq
k\leq N\atop k \neq j}\frac{1}{x_{jk}}\right).\eeqa

As observed in \cite{ShastrySutherland}, the term
$1-\theta_{ij}\theta^{\dagger}_{ij}$ is simply a fermionic
exchange operator
\begin{equation}
\kappa_{ij}\equiv 1-\theta_{ij}\theta^{\dagger}_{ij}=
1-(\theta_{i}-\theta_j)(\partial_{\theta_i}-\partial_{\theta_j})
\, ,
\end{equation}
whose action on a function
     $ f(\theta_i,\theta_j) $ is
\begin{equation}
\kappa_{ij}\, f(\theta_i,\theta_j)= f(\theta_j,\theta_i)\,
\kappa_{ij}\, .
\end{equation}
This provides a substantial simplification in that the model can
be studied using the exchange-operator formalism
\cite{Polychronakos:1992zk}.

We now remove  the contribution of the ground-state wave function
$\psi_0$ from the Hamiltonian $\mc{H}$.  The unique square
integrable function satisfying $Q\psi_0=0$ is
\begin{equation}
\psi_0(x) =\prod_{1\leq j<k\leq N}\left( x_{jk}\right)^\beta
e^{-\frac12 \omega\|x\|^2 } , \qquad\|x\|^2=\sum_{i=1}^N x_j^2,
\end{equation}
with $\om>0$ and $\beta>-1/2$.\footnote{ When the coupling
constant $\beta\leq-1/2$, the supersymmetry is spontaneously
broken \cite{Freedman}. In this article, only the supersymmetric
phase is considered ($\beta>-1/2$).}  We restrict $\beta$ to be a
positive integer: odd or even for antisymmetric (fermionic) or
symmetric (bosonic) ground state respectively.   The transformed
(or gauged) Hamiltonian
\begin{equation}
\bar{\mc{H}}\equiv
     \psi_0(x)^{-1}\, \mc{H}\, \psi_0(x)
\, ,
\end{equation}
    reads\footnote{From now on, $\sum_i$ means $\sum_{i=1}^N$,
    $\sum_{i<j}$ means $\sum_{1\leq i< j\leq N}$, $\sum_{j\neq i}$
    means $\sum_{1\leq j\leq N\atop j\neq i}$, etc.}
\begin{equation} \label{shjack}
\bar{\mc{H}}= 2\om\sum_i (x_i
\partial_i+\theta_i\partial_{\theta_i})- \sum_i
\partial_i^2-2\beta \sum_{i<j}\frac{ 1} {x_{ij}}(
\partial_i-\partial_j)+2\beta\sum_{i<j}\frac{1}{x_{ij}^2}(1-\kappa_{ij})
\, ,
\end{equation}
where $\partial_i= \partial_{x_i}$. The generalized Hermite
polynomials in superspace will be orthogonal eigenfunctions of
this Hamiltonian.

The supersymmetric Hamiltonian $\bar{\mc{H}}$ is self-adjoint with
respect to the physical scalar product given by
  \beq \label{defscaprod}\langle
F(x,\theta),G(x,\theta)\rangle_{\beta,\omega}=\prod_{j}\left(\int_{-\infty}^\infty
dx_j\int d\theta_j\theta_j\right) \prod_{k\leq
l}(x_{kl})^{2\beta}e^{-\omega\| x \|^2}
F(x,\theta)^*G(x,\theta),\eeq where $F$ and $G$ are arbitrary
functions.  The complex conjugation $*$  is defined such
that\beq\label{defcomplex}
(\theta_{i_1}\cdots\theta_{i_m})^*\theta_{i_1}\cdots\theta_{i_m}=1\qquad\mbox{and}\qquad
x_j^*=x_j.\eeq In other words, $\theta_j^*$ behaves  as
$\theta_j^\dagger=\partial_{\theta_j}$.  The integration over the
Grassmannian variable is the standard Berezin integration, i.e.,
\beq \int d\theta=0,\qquad\int d\theta \,\theta=1.\eeq

\subsection{Symmetric superfunctions}

The Hamiltonian  $\bar{\mc{H}}$ leaves invariant the space of
symmetric superpolynomials, $P^{S_N}$, invariant under the
simultaneous action of $ \kappa_{ij} $  and  $ K_{ij} $, where  $
K_{ij} $ is the exchange operator acting on the  $ x_i $
variables:
\begin{equation}
            K_{ij}f(x_i, x_j)=f(x_j, x_i)K_{ij} \, .
\end{equation} A polynomial $f$ thus belongs to $P^{S_N}$ if it is
invariant under the action of the product
   \beq \mc{K}_{ij}= \kappa_{ij}K_{ij} \, , \eeq
that is, if $\mc{K}_{ij} f=f$ for all $i,j$.
Con
As argued in \cite{DLM1},  superpartitions provide a labelling of
symmetric superpolynomials. We recall that a superpartition
$\Lambda$ in the  $ m $-fermion sector is a sequence of integers
composed of two standard partitions separated by a semicolon:
$\Lambda=(\Lambda^a ;\Lambda^s)$. The first partition,
$\Lambda^a$, is associated to an antisymmetric function and has
thus distinct parts.  While the second one, $\Lambda^s$, is
associated to a symmetric function and is a standard partition. To
be more specific,

\begin{definition}\label{dessuperpart} Let $\La$ be a sequence of $N$
integers $\La_j\geq 0$ that decomposes into two parts
separated by a semi-colon :
  \begin{equation}
\Lambda=(\Lambda_1,\ldots,\Lambda_m;\Lambda_{m+1},\ldots,\Lambda_{N})\equiv
(\Lambda^a ; \Lambda^s). \end{equation} Then,  $\La$ is a \emph{
superpartition} iff $\Lambda^a=(\Lambda_1,\ldots,\Lambda_m)$
{with} $ \Lambda_i>\Lambda_{i+1}\geq 0$ for $ i=1, \ldots m-1$ and
$\Lambda^s= (\Lambda_{m+1},\ldots,\Lambda_{N})$ \mbox{with} $
\Lambda_j \ge \Lambda_{j+1}\geq 0 $ for $j=m+1,\dots,N-1$.
\end{definition}

Notice that in the zero-fermion sector, the semicolon is usually
omitted and
     $ \Lambda $  reduces then to  $ \Lambda^s $. We denote the degree of a
superpartition and its fermionic number respectively by:
\begin{equation}
|\Lambda|=\sum_{i=1}^{N}\Lambda_i,\qquad\mbox{and}\qquad
\overline{\underline{\Lambda}} = m\, .
\end{equation}

To every superpartition $\Lambda$, we associate a unique
composition $\Lambda_c$ obtained by deleting the semi-colon ,
i.e., $\Lambda_c=(\Lambda_1,\ldots,\Lambda_N)$.  Finally, the
partition rearrangement in non-increasing order of the entries of
$\Lambda$ is denoted $\Lambda^*$.

\medskip

We now introduce three bases for the space of symmetric
superfunctions.

\subsection{Monomial symmetric superfunctions  }

The monomial symmetric superfunctions \cite{DLM1}, superanalogues
to the monomial symmetric functions, are defined as follows:
\begin{equation}
m_{\Lambda}(x,\theta)=\sm{\Lambda_1,\ldots,
\Lambda_m;\Lambda_{m+1},\ldots,\Lambda_{N}}(x,\theta)={\sum_{\sigma\in
S_{N}}}' \theta^{\sigma(1, \ldots, m)}x^{\sigma(\Lambda)},
\end{equation} where the prime indicates that the  summation is
restricted to distinct terms, and where
\begin{equation}
x^{\sigma(\Lambda)}=x_1^{\Lambda_{\sigma(1)}} \cdots
x_m^{\Lambda_{\sigma(m)}} x_{m+1}^{\Lambda_{\sigma(m+1)}} \cdots
x_{N}^{\Lambda_{\sigma(N)}}  \quad {\rm{and}} \quad
\theta^{\sigma(1, \ldots, m)} = \theta_{\sigma(1)} \cdots
\theta_{\sigma(m)} \, .
\end{equation}

For technical manipulations, it is convenient to reexpress the
symmetric group action on the variables using exchange operators.
To a reduced decomposition $\sigma =
    \sigma_{i_1} \cdots
    \sigma_{i_n}$ of an element $\sigma$ of the symmetric group $S_N$,
we associate
    $\mc{K}_{\sigma}$, which stands for $\mc{K}_{i_1,i_1+1} \cdots
\mc{K}_{i_n,i_n+1}$.
    The monomial superfunction $m_{\Lambda}$ can thus be rewritten
    as \begin{equation} \label{monod}
    m_{\Lambda}= \frac{1}{f_{\Lambda}} \sum_{\sigma \in S_N} \mc{K}_{\sigma}
    \left( \theta_1 \cdots \theta_m x^{\Lambda}\right) \, ,\qquad \quad
f_{\Lambda} =f_{\Lambda^s}= n_{\Lambda^s}(0)!\, n_{\Lambda^s}(1)!
\,
    n_{\Lambda^s}(2) ! \cdots \, ,
\end{equation}
    with  $n_{\Lambda^s}(i)$ the number of $i$'s in $\Lambda^s$,
    the symmetric part of $\Lambda=(\Lambda^a;\Lambda^s)$.

When the bosonic variables lay on a unit circle in the complex
plane, that is, $x_j\in\mathbb{C}$ and $x_j^*=1/x_j$ for all $j$,
the monomials are orthogonal with respect to the following scalar
product : \beq \langle
A(x,\theta),B(x,\theta)\rangle=\prod_{j}\left(\frac{1}{2\pi i}
\oint \frac{ dx_j}{x_j}\int d\theta_j\,\theta_j\right)
\left[A(x,\theta)^* B(x,\theta)\right] \, ,\eeq where the complex
conjugation of the fermionic variables was defined in
(\ref{defcomplex}).

\subsection{Jack  superpolynomials}

The generators of the second basis are the superanalogues of the
Jack polynomials first introduced in \cite{DLM1} and whose
definition has been clarified in \cite{DLM2} and \cite{DLM3} (see
also \cite{DLM4} for a pedagogical presentation). The Jack
superpolynomials $J_\Lambda$ are defined to be the unique
orthogonal eigenfunctions of the (gauged) stCMS Hamiltonian
$\mc{H}_2$ that decompose triangularly (with respect to a given
ordering) in terms of supermonomials.

To clarify this definition we need to introduce a partial ordering
on superpartitions and a scalar product.  We first recall the
usual dominance ordering on two partitions $\lambda$ and $\mu$ of
the same degree: $\lambda \leq \mu$ iff
${\lambda}_1+\ldots+{\lambda}_k\leq{\mu}_1+\ldots+{\mu}_k$ for all
$k$. The partial ordering on superpartitions is similar.

\begin{definition}\label{defdominance} The \emph{dominance ordering}
$\leq$ is  such
that $\Omega\leq\Lambda$  if either $\Omega^*<\Lambda^*$, or
$\Omega^*=\Lambda^*$ and
${\Omega}_1+\ldots+{\Omega}_k\leq{\Lambda}_1+\ldots{\Lambda}_k$
for all $k$.
\end{definition}
\n Note that this ordering is simply the usual ordering on
compositions \cite{BF2}.

Next, we introduce the physical scalar product of the stCMS model
: \beq \langle
A(x,\theta),B(x,\theta)\rangle_{\beta}=\prod_{j}\left(\frac{1}{2\pi
i} \oint \frac{ dx_j}{x_j}\int
d\theta_j\,\theta_j\right)\left[\prod_{k\neq
l}\left(1-\frac{x_k}{x_l}\right)^\beta
A(x,\theta)^*B(x,\theta)\right],\eeq where the variable $x_j$
represents  the $j^{\mbox{\tiny th}}$ particle's position on the
unit circle in the  complex plane; this means $x_j^*=1/x_j$.  We
recall again that the complex conjugation of the fermionic
variables was defined in (\ref{defcomplex}).

The simplest definition of the Jack polynomial in  superspace is
the following.
\begin{definition} \cite{DLM3} The \emph{Jack polynomials}
$J_\Lambda\equiv J_\La(x,\theta;1/\beta)$ in superspace are the
unique functions in $P^{S_N}$ such that
  \beq\label{defjack1}
   \langle
J_\Lambda,J_\Omega\rangle_{\beta}\propto \delta_{\Lambda,\Omega}
\qquad\mbox{and}\qquad
J_\Lambda=m_\Lambda+\sum_{\Omega<\Lambda}t_{\Lambda,\Omega}(\beta)m_\Omega.
\eeq
\end{definition}

\subsection{Power-sum superpolynomials}

The space of superpolynomials is generated by $2N$ algebraically
independent variables: $N$ commutative $x_i$'s and $N$
anticommutative $\theta_i$'s.  Equivalently, $2N$ independent and
symmetric quantities can generate the space $P^{S_N}$ of
superpolynomials invariant under the $\mc{K}_{ij}$'s.  An obvious
choice for the $2N$ independent quantities is the following
\cite{DLM4}: \beq p_n=\sum_ix_i^n=m_{(;n)}\qquad\mbox{and}\qquad
q_{n-1}=\sum_i\theta_ix_i^{n-1}=m_{(n-1;0)},\eeq where
$n=1,\ldots,N$.  The $p_n$'s are the standard power sums while the
$q_{n-1}$'s are their fermionic counterparts.

The power-sum basis in superspace, denoted $\{p_\La\}_\La$, is
constructed from the generalized product of power sums: \beq
p_\La=q_{\La_1}\cdots q_{\La_m}p_{\La_{m+1}}\cdots
p_{\La_N},\qquad m=\overline{\underline{\La}}.\eeq

\medskip

In the next section, we generalize (i.e., deform via the parameter
$\om$) the three preceding bases of $P^{S_N}$; they will thus
become bases of eigenfunctions for the srCMS Hamiltonian
$\bar{\mc{H}}$.

\section{Conserved quantities}

The simplest way to define the srCMS conserved operators is from
the stCMS's ones using a mapping from the trigonometric to the
rational case. Trigonometric conserved quantities are constructed
using Dunkl operators, denoted $\mc{D}_j$,  and a projection onto
the space of symmetric superpolynomials, denoted
$\,\big|_{P^{S_N}}$.

\subsection{Trigonometric Case}
The trigonometric Dunkl operators are the following differential
operators \cite{Cherednik, Bernard}: \footnote{These operators are
also called Cherednik operators and denoted $\xi_j$ by some
authors (cf. \cite{BF2}).} \beq\label{defchered}
\D_j=x_j\partial_j+\beta\sum_{k<j}\frac{x_j}{x_{jk}}(1-K_{jk})+\beta\sum_{k>j}\frac{x_k}{x_{jk}}(1-K_{jk})-\beta(j-1)\qquad
j=1,\ldots,N.\eeq
  They  satisfy the
degenerate Hecke relations: \beq \label{hecke} K_{i,i+1} {\mathcal
D}_{i+1} - {\mathcal D}_i K_{i,i+1} = \beta \qquad {\rm and}
\qquad K_{j,j+1} {\mathcal D}_{i} = {\mathcal D}_i K_{j,j+1} \quad
       (i \neq j,j+1) \, .
\eeq The $4N$  conserved charges are obtained via the projection
of the following quantities: \beq\begin{array}{lll} {\mathcal H}_n
&=& \ds\sum_{i=1}^N
       {\mathcal D}_i^n ,\\
{\mathcal Q}_{n-1}&=& \ds\sum_{w \in S_N} {\mathcal K}_{w}
\theta_1 {\mathcal D}_1^{n-1}  ,\\
  {{\mathcal Q}^\dagger_{n-1}} &=&
\ds\sum_{w \in S_N} {\mathcal K}_{w}
  \frac{\d}{ \d \theta_1} {\mathcal D}_1^{n-1} ,\\
{\mathcal I}_{n-1}&=& \ds\sum_{w \in S_N} {\mathcal K}_{w}
\theta_1 \frac{\d}{ \d \theta_1} {\mathcal D}_1^{n-1} ,\end{array}
\eeq where $n=1,2,\ldots,N$. Specially, upon projection onto
$P^{S_N}$ (that is, when limiting their action to functions
belonging to $P^{S_N}$), they become $4N$ conserved operators of
the stCMS model. However, only the $2N$ bosonic ones are in
involution: \beq
[\mc{H}_m|_{P^{S_N}},\mc{H}_n|_{P^{S_N}}]=[\mc{H}_m|_{P^{S_N}},\mc{I}_{n-1}|_{P^{S_N}}]=[\mc{I}_{m-1}|_{P^{S_N}},\mc{I}_{n-1}|_{P^{S_N}}]=0.\eeq

\subsection{Rational case}
  We now define the rational Dunkl operators (without the ground state
contribution) \cite{Dunkl}:
\begin{equation} \label{defDi} D_j=\frac{\partial}{\partial
x_j}+\beta\sum_{k\neq j}\frac{1}{x_{jk}}(1-K_{jk})\, .
\end{equation}
They satisfy \beq [D_i,D_j]=0\qquad\mbox{and}\qquad
K_{ij}D_i=D_jK_{ij}.\eeq
  These operators are not self-adjoint with
respect to the physical scalar product (\ref{defscaprod}). The
Hermitian adjoint of $D_j$ is
\begin{equation} \label{Ddag}D_j^\dagger=2\omega x_j-D_j\, . \end{equation}
Note that $K_{ij}^{\dagger}=K_{ij}$ since the measure in
(\ref{defscaprod}) is $S_N$-invariant. We can construct a set of
$N$ commuting and self-adjoint operators as follows:\footnote{In
essence, $\mc{D}_j^\om$ corresponds  to the operator $d_j$ defined
in \cite{UWa4},  plus the exchange term $\beta\sum_{k>j}K_{jk}$.
This deformation was inspired by \cite{Bernard}.} \beq
\D_j^\om=\frac{1}{2\om}D_j^\dagger
D_j+\beta\sum_{k>j}K_{jk}-\beta(N-1)=\frac{1}{2\om}D_j
D_j^\dagger-\beta\sum_{k<j}K_{jk}-\beta(N-1)-1,\eeq for
$j=1,\ldots, N$.   The importance of the $\D_j^\om$'s becomes more
transparent  if we rewrite them as \cite{BF2} \beq
\D_j^\om=\D_j-\frac{1}{2\om}D_j^2\eeq (we stress that on the
r.h.s., there is a $\D_j$ defined in \eqref{defchered} and a
$D_j$) which implies
  \beq \lim_{\om\rightarrow \infty}\D_j^\om=\D_j.\eeq
This result suggests that we can construct $2N$ Hermitian and
commuting quantities by simply `$\om$-deforming' the set
$\{\mc{H}_n,\mc{I}_{n-1}\}$.  We thus define
\beq\label{defDom}\begin{array}{lll}
\mc{H}_n^\omega&=&\ds\sum_j\left(\D_j^\om\right)^n,\\
\mc{I}_{n-1}^\om&=&\ds\sum_{w \in S_N} {\mathcal
K}_{w}\theta_1\partial_{\theta_1}\left(\D_1^\om\right)^{n-1}.\end{array}\eeq

\begin{lemma}\label{lemmasimtrans}The rational operators
$\{\mc{H}_n^\omega, \mc{I}_{n-1}^\omega\}$ are related to their
trigonometric counterparts  $\{\mc{H}_n, \mc{I}_{n-1}\}$ by the
following similarity transformations :
\beq\label{simtrans}\mc{H}_n^\omega
=e^{-\frac{1}{4\om}\Delta}\,\mc{H}_n\,
e^{\frac{1}{4\om}\Delta}\qquad\mbox{and}\qquad
\mc{I}_{n-1}^\omega=e^{-\frac{1}{4\om}\Delta}\,\mc{I}_{n-1}\,
e^{\frac{1}{4\om}\Delta},\eeq where $ n=1,\ldots, N$ and \beq
\Delta=\sum_j D_j^2.\eeq
\end{lemma}
  \n {\it Proof}.  Since $D_j$ and $\Delta$ commute, we first remark
that \cite{BF2}
  \beq [\D_j^\om,\Delta]=[\D_j,\Delta]=-2D_j^2 \, ,\eeq
  Then, using the
Baker-Campbell-Hausdorff formula, we get
  \beq
  e^{-\frac{1}{4\om}\Delta}\,
  \D_j\,e^{\frac{1}{4\om}\Delta}=\D_j-\frac{1}{4\om}[\Delta,\D_j]+
  \frac{1}{2!}\frac{1}{(4\om)^2}[\Delta,[\Delta,\D_j]]\,-\,\cdots
=\D_j-\frac{1}{2\om}D_j^2=\D_j^\om,\eeq which implies the relation
(\ref{simtrans}). \hfill $\square$

The operator $\Delta$ preserves the space $P^{S_N}$ of symmetric
superpolynomials. Therefore, the similarity transformation is not
affected by the projection operation onto the space of symmetric
superpolynomials. More explicitly, if $F(\D_j)$ and $G(\D_j)$ are
operators  satisfying
  $\mc{K}_{ij}F=F$ and $\mc{K}_{ij}G=G$, then
\beq e^{-\frac{1}{4\om}\Delta}F
e^{\frac{1}{4\om}\Delta}\big|_{P^{S_n}}
=e^{-\frac{1}{4\om}\Delta}\big|_{P^{S_n}}\,F\big|_{P^{S_n}}\,
e^{\frac{1}{4\om}\Delta}\big|_{P^{S_n}},\eeq and therefore, \beq
[F,G]\big|_{P^{S_N}}=0\qquad\Longrightarrow\qquad\left[
  e^{-\frac{1}{4\om}\Delta}F
e^{\frac{1}{4\om}\Delta}\big|_{P^{S_n}}\, , \,
e^{-\frac{1}{4\om}\Delta}G
e^{\frac{1}{4\om}\Delta}\big|_{P^{S_n}}\right]=0 .\eeq

  This observation and
Lemma \ref{lemmasimtrans} prove the following proposition.

\begin{proposition}\label{quantcons} For $n, m\in\{1,2,\ldots,N\}$, the
set of operators $\{\mc{H}_n^\om,\mc{I}_{n-1}^\om\}$ defined by
equations (\ref{defDom}) or (\ref{simtrans}) is such that \beq
[\mc{H}_m^\om|_{P^{S_N}},\mc{H}_n^\om|_{P^{S_N}}]=[\mc{H}_m^\om|_{P^{S_N}},\mc{I}_{n-1}^\om
|_{P^{S_N}}]=[\mc{I}_{m-1}^\om|_{P^{S_N}},\mc{I}_{n-1}^\om|_{P^{S_N}}]=0.\eeq
\end{proposition}

We stress that the supersymmetric Hamiltonian defined in
(\ref{shjack}) can be rewritten as \beq \bar{\mc{H}}=2\om
\mc{H}^\om_1|_{P^{S_N}}+2\om \mc{I}^\om_0|_{P^{S_N}}.\eeq
Furthermore, the fermionic operators
\beq\mc{Q}_1^\om|_{P^{S_N}}=e^{-\frac{1}{4\om}\Delta}\mc{Q}_1
e^{\frac{1}{4\om}\Delta}|_{P^{S_N}}\qquad\mbox{and}\qquad
{\mc{Q}_1^\om}^\dagger|_{P^{S_N}}=e^{-\frac{1}{4\om}\Delta}{\mc{Q}_1}^\dagger
e^{\frac{1}{4\om}\Delta}|_{P^{S_N}}\eeq are essentially the
supersymmetric generators (\ref{fcharges}) without the ground
state contribution.

As we will see in the next section,  the generalized Hermite
superpolynomials will be orthogonal eigenfunctions of the $2N$
quantities $\mc{H}^\om_n|_{P^{S_N}}$ and
$\mc{I}_{n-1}^\om|_{P^{S_N}}$, $n=1,\ldots, N$.

\section{From Jack to Hermite polynomials in superspace}

\subsection{Algebraic construction and triangularity}
  The bijection established in (\ref{simtrans}) between
the srCMS and stCMS conserved quantities allows us to construct
generalized Hermite superpolynomials directly from the Jack
superpolynomials. The Jack polynomials in superspace are the
common eigenfunctions of the $\mc{H}_n$'s and $\mc{I}_{n-1}$'s,
that is, \beq \mc{H}_n
J_\Lambda=\varepsilon_{\Lambda,n}J_\Lambda\qquad\mbox{and}\qquad
\mc{I}_{n-1} J_\Lambda=\epsilon_{\Lambda,n-1}J_\Lambda,\eeq where
the eigenvalues are such that \beq
\label{linepsilon}\lim_{\beta\rightarrow
0}\varepsilon_{\Lambda,n}=\sum_{i=1}^N\Lambda_i^n\qquad\mbox{and}\qquad\lim_{\beta\rightarrow
0}\epsilon_{\Lambda,n-1}=\sum_{i=1}^{\underline{\overline{\Lambda}}}\Lambda_i^{n-1}
. \eeq Note the following two special cases \beq
\varepsilon_{\Lambda,1}=|\Lambda|
\qquad\mbox{and}\qquad\epsilon_{\Lambda,0}=\underline{\overline{\Lambda}}\,.\eeq
Given that the stCMS and srCMS conserved operators are related by
a similarity transformation, i.e., \beq
\mc{H}_{n}=e^{\frac{1}{4\om}\Delta}\mc{H}_n^\om
e^{-\frac{1}{4\om}\Delta}
\qquad\mbox{and}\qquad\mc{I}_{n-1}=e^{\frac{1}{4\om}\Delta}\mc{I}_{n-1}^\om
e^{-\frac{1}{4\om}\Delta},\eeq the eigenfunctions of the srCMS
charges, defined  by \beq \label{eigenvaluehermite}\mc{H}_n^\om
J_\Lambda^\om=\varepsilon_{\Lambda,n}J_\Lambda^\om\qquad\mbox{and}\qquad
\mc{I}_{n-1}^\om
J_\Lambda^\om=\epsilon_{\Lambda,n-1}J_\Lambda^\om,\eeq are
directly found to be
  \beq\label{defhermite1}
J_\Lambda^\om\equiv
J_\La(x,\theta;1/\beta,\om)=e^{-\frac{1}{4\om}\Delta}J_\Lambda(x,\theta;1/\beta).
\eeq
\begin{definition}
The generalized Hermite polynomials in superspace are given by
\begin{equation}
J_\Lambda^\om=e^{-\frac{1}{4\om}\Delta}J_\Lambda \, .
\end{equation}
\end{definition}

We stress that this definition makes sense since, as it is proved
in \cite{DLM5}, the action of the operator $\Delta$ on monomials
is finite and triangular when $|\Lambda|<\infty$ and $N<\infty$.
Therefore, the relation between the $J_\Lambda^\om$'s and  the
$J_\Lambda$'s is bijective. More precisely, $\Delta$ has degree
$-2$ in $x$ and $0$ in $\theta$. Given that $\Delta m_\La$ or
$\Delta J_\La$ has a finite polynomial decomposition, we may write
\beq \label{triH} J_\Lambda^\om=
e^{-\frac{1}{4\om}\Delta}J_\Lambda=J_\La+\sum_{\Omega<_u\La}A_{\La
\Om}(\beta,\om,N)J_\Om,\eeq where the ordering $\leq_u$ is defined
as follows.

\begin{definition}\label{defuorder}
The \emph{$u$-ordering} is  such that $\Om\leq_u\La$ if either
$\Om=\La$, or $|\Om|=|\La|-2n$ for $n=1,2,3,\ldots$
\end{definition}

\n The action of the rational conserved charges on the Jack
superpolynomials is thus triangular.

\begin{lemma} \label{triangn}
The $\mc{H}_n^\om$'s and the $\mc{I}_{n-1}^\om$'s act triangularly
on the $J_\La$'s: \beq \mc{H}^\om_n
J_\La=\varepsilon_{\La,n}J_\La+\sum_{\Om<_u\La}B_{\La
\Om}J_\Om\qquad\mbox{and}\qquad\mc{I}^\om_{n-1}
J_\La=\epsilon_{\La,n-1}J_\La+\sum_{\Om<_u\La}C_{\La
\Om}J_\Om,\eeq where $B_{\La \Om}$ and $C_{\La \Om}$ are some
coefficients in $\beta$, $\om$ and $N$.
\end{lemma}

\subsection{Orthogonality}

The triangular action of the conserved operators on the Jack
superpolynomial  basis and their  hermiticity  are essential
properties : they imply the orthogonality of the generalized
Hermite superpolynomials.
\begin{theorem}\label{theodefhermite2}
The generalized Hermite polynomials, $J^{\om}_{\Lambda}$, form the
unique basis of $P^{S_N}$ such that
  \beq \label{defhermite2}
\langle J_\Lambda^\om,J_\Om^\om\rangle_{\beta,\omega} \propto
\delta_{\La,\Om}\qquad{and}\qquad
J_\La^\om=J_\La+\sum_{\Om<_u\La}w_{\La
\Om}(\beta,\om,N)\,J_\Om.\eeq
\end{theorem}
\n {\it Proof.} The generalized Hermite superpolynomials are
triangular and monic (see \eqref{triH}).  To show their
orthogonality, we simply need to see that they are eigenfunctions
(see \eqref{eigenvaluehermite}) of the complete set of
self-adjoint operators $\{ \mc{H}_n^\om ,\mc{I}_{n-1}^\om\}$.  The
completeness follows from (\ref{linepsilon}) which implies that
  the set
\beq
\big\{(\varepsilon_{\La,1},\varepsilon_{\La,2},\ldots,\varepsilon_{\La,N};\epsilon_{\La,0},\epsilon_{\La,1},\ldots,\epsilon_{\La,N-1})\quad
\big|\quad
|\Lambda|=n,\quad\underline{\overline{\Lambda}}=m\big\}\eeq does
not contains identical multiplets (when $\beta$ is viewed as a formal
parameter).  There is uniqueness because the Gram-Schmidt
orthonormalization procedure ensures that there is at most one
orthonormal family with a given triangularity\footnote{The fact that our
order is partial makes the condition stronger.  That is, given a partial
order, it is not sure that an orthonormal basis triangular with respect
to that order exist.  But if it exists, it is unique.}. \hfill$\square$

Theorem \ref{theodefhermite2} can be interpreted as another
definition of the generalized Hermite polynomials in superspace.
The following theorem gives a third definition of the generalized
Hermite polynomials: as  triangular eigenfunctions of the
supersymmetric Hamiltonian $\mc{H}_1^\om$.

\begin{theorem}\label{theodefhermite3}
The generalized Hermite polynomials, $J^{\om}_{\Lambda}$, form the
unique basis of $P^{S_N}$ such that
  \beq \label{defhermite3}
\mc{H}_1^\om J_\Lambda^\om=|\La|J_\La^\om\qquad{and}\qquad
J_\La^\om=J_\La+\sum_{\Om<_u\La}w_{\La
\Om}(\beta,\om,N)\,J_\Om.\eeq
\end{theorem}
\n{\it Proof.}  The $J_\Lambda^\om$'s are triangular and
eigenfunctions of $\mc{H}_1^\om$ with the right eigenvalue (see
\eqref{triH} and \eqref{eigenvaluehermite}).  Uniqueness is proved
as follows. We suppose the existence of $\tilde{J}_\La^\om$
satisfying equation (\ref{defhermite3}), so that \beq
J_\La^\om-\tilde{J}_\La^\om=\sum_{\Om<_u\La}D_{\La
\Om}J_\Om=\sum_{i=1}^n D_{\La \Om^{(i)}}J_{\Om^{(i)}}, \eeq where
$\Om^{(1)}\preceq_u \ldots\preceq_u \Om^{(n)}<\La$ and $\preceq_u$
stands for a total ordering compatible with $\leq_u$. On the one
hand, \beq \label{unique1}
\mc{H}_1^\om(J_\La^\om-\tilde{J}_\La^\om)= \varepsilon_{\La,1}
(J_\La^\om-\tilde{J}_\La^\om)
=\varepsilon_{\La,1}\sum_{i=1}^nD_{\La,\Om^{(i)}}J_{\Om^{(i)}}
=\varepsilon_{\La,1}D_{\La \Om^{(n)}}J_{\Om^{(n)}}+\ldots, \eeq
where `$\ldots$'  represents lower terms with respect to the
ordering $\preceq_u$.  On the other hand, using Lemma~\ref{triangn}, we
get \beq \label{unique2}
\mc{H}_1^\om(J_\La^\om-\tilde{J}_\La^\om)=\sum_{i=1}^n D_{\La
\Om^{(i)}}\left(\varepsilon_{\Om^{(i)},1}J_{\Om^{(i)}}
+\sum_{\Gamma<_u\Om^{(i)}}E_{\Om^{(i)}\Gamma}J_\Gamma\right)
=\varepsilon_{\Om^{(n)},1}D_{\La
\Om^{(n)}}J_{\Om^{(n)}}+\ldots\eeq
   Equations (\ref{unique1}) and (\ref{unique2}) imply
$\varepsilon_{\La,1}=\varepsilon_{\Om^{(n)},1}$. However, this is
impossible since $\varepsilon_{\La,1}=|\La|$ and
$\varepsilon_{\Om^{(n)},1}=|\Om^{(n)}|<|\La|$.  The function
$J_\La^\om$ is thus unique. \hfill$\square$

\section{Maximal superintegrability}

As an offshoot of our analysis of the srCMS conserved quantities,
we will now  demonstrate  that this model is superintegrable.  A
bosonic Hamiltonian model with $N$ variables $x_i$ is
superintegrable if it possesses more than $N$ functionally
independent conserved quantities.   The maximum number of such
quantities is $2N-1$ while only $N$ can be in involution at the
same time (cf. \cite{Kuznetsov}).  A fermionic extension of a
model that contains also $N$ Grassmannian variables $\theta_i$ is
superintegrable if it has more than $2N$ conserved and independent
operators.  It is maximally superintegrable if the total number of
operators that commute with the Hamiltonian is $4N-2$.

The supersymmetric generalization of the rational CMS model
without harmonic confinement ($\om=0$) is maximally
superintegrable.  This was shown in \cite{DLM4} using the Dunkl-
operator formalism.  When $\om\neq 0$, the superintegrability is
even  simpler to prove.  It uses a bijection between the srCMS
model and the free supersymmetric model, whose  Hamiltonian is
\beq \mathrm{H}=\{\mathrm{Q},\mathrm{Q}^\dagger\}=\sum_j
x_j\partial_j=\mc{H}_1,\eeq with the fermionic charges formally
defined as \beq \mathrm{Q}=\sum_j
\theta_j(x_j\partial_j)^{1/2}\qquad\mbox{and}\qquad
\mathrm{Q}^\dagger=\sum_j
\partial_{\theta_j}(x_j\partial_j)^{1/2}.\eeq
The conserved and independent operators are simply
\beq\label{eqfreeconsquant}\begin{array}{lll}\mathrm{H}_n=\ds\sum_j
(x_j\partial_j)^n ,& \mathrm{I}_{n-1}=\ds\sum_j
\theta_j\partial_{\theta_j}(x_j\partial_j)^{n-1}
,&n=1,\ldots,N,\cr
\mathrm{J}_{n}=\ds\mathrm{H_n}\mathrm{L}_0-\mathrm{H}_1\mathrm{L}_{n-1},&
\mathrm{K}_{n-1}=\ds\mathrm{I_n}\mathrm{M}_0-
\mathrm{I}_1\mathrm{M}_{n-1},&n=1,\ldots,N-1,
\end{array}\eeq where $\mathrm{L}_n$ and $\mathrm{M}_n$  are defined
as \beq \mathrm{L}_{n-2}=\sum_j\ln
|x_j|(x_j\partial_j)^{n-1}\qquad\mbox{and}\qquad
\mathrm{M}_{n-2}=\sum_j\ln
|x_j|\theta_j\partial_{\theta_j}(x_j\partial_j)^{n-1},\eeq if
$n\geq 1$. The commutativity of the $\mathrm{H}_n$'s and the
$\mathrm{I}_{n-1}$'s is immediate:\beq
[\mathrm{H}_m,\mathrm{H}_n]=[\mathrm{H}_m,\mathrm{I}_{n-1}]=[\mathrm{I}_{m-1},\mathrm{I}_{n-1}]=0.\eeq
The conservation (commutation with $\mathrm{H}\equiv\mathrm{H}_1$)
of the $\mathrm{J}_n$'s and $\mathrm{K}_{n-1}$'s is an obvious
consequence of the following commutation relations: \beq
[\mathrm{H}_n,\mathrm{L}_m]=n\mathrm{H}_{n+m},\qquad
[\mathrm{H}_n,\mathrm{M}_m]=n\mathrm{I}_{n+m}.\eeq Notice that the
$\mathrm{L}_n$'s generate a Virasoro algebra (with zero central
charge), i.e., \beq
[\mathrm{L}_n,\mathrm{L}_m]=(n-m)\mathrm{L}_{n+m}.\eeq

As we showed in the beginning of  Section 3, the srCMS Hamiltonian
$\mc{H}_1^\om$ and the free Hamiltonian
$\mathrm{H}=\mathrm{H}_1=\mc{H}_1$ (recall that $\mc{H}_1$ denotes
the first trigonometric conserved quantity) are related by a
similarity transformation:
\beq\label{eqsimtransfree}\mc{H}_1^\om=e^{-\frac{1}{4\om}\Delta}\mathrm{H}_1
e^{\frac{1}{4\om}\Delta}\big|_{P^{S_N}}=e^{-\frac{1}{4\om}\Delta}\mathcal{H}_1
e^{\frac{1}{4\om}\Delta}\big|_{P^{S_N}}.\eeq This means in
particular that, instead of $\om$-deforming  the trigonometric
conserved quantities $\mc{H}_n$ to obtain the Hamiltonians
$\mc{H}_n^\om$ of the srCMS model (cf. Section 3.1), one can
construct another set of rational conserved operators
$\mathrm{H}_n^\om$ by transforming the free ones.\footnote{There
is no algebraic relation between the sets $\{\mc{H}_n^\om,
\mc{I}_{n-1}^\om\}$  and $\{\mathrm{H}_n^\om,
\mathrm{I}_{n-1}^\om\}$, obtained
  from the trigonometric and the free model respectively.}  More generally,
equations (\ref{eqsimtransfree}) and (\ref{eqfreeconsquant}) give
an algebraic construction of all srCMS conserved operators.

\begin{proposition}
The srCMS model is maximally superintegrable; an independent set
of  conserved operators can be constructed as follows:
\beq\label{eqsimtransfree2}\begin{array}{llcll}
\mathrm{H}_n^\om\big|_{P^{S_N}}&= \ds
e^{-\frac{1}{4\om}\Delta}\,\mathrm{H}_n\,
e^{\frac{1}{4\om}\Delta}\big|_{P^{S_N}},&\qquad&
\mathrm{I}_{n-1}^\om\big|_{P^{S_N}}&=\ds
e^{-\frac{1}{4\om}\Delta}\,\mathrm{I}_{n-1}\,
e^{\frac{1}{4\om}\Delta}\big|_{P^{S_N}},\cr
\mathrm{J}_{n}^\om\big|_{P^{S_N}}&=\ds
e^{-\frac{1}{4\om}\Delta}\,\mathrm{J}_{n}\,
e^{\frac{1}{4\om}\Delta}\big|_{P^{S_N}},&\qquad&
\mathrm{K}_{n-1}^\om\big|_{P^{S_N}}&=\ds
e^{-\frac{1}{4\om}\Delta}\,\mathrm{K}_{n-1}\,
e^{\frac{1}{4\om}\Delta}\big|_{P^{S_N}}.\end{array}\eeq
  \end{proposition}

\section{Other bases of eigenfunctions}

The preceding construction of commuting operators
  in the srCMS model
furnishes directly two other sets of eigenfunctions.

  Indeed, let
the superpolynomial  $m_\La^\om\equiv m_\La(x,\theta;\om,\beta)$
and the power-sum product $p_\La^\om\equiv
p_\La(x,\theta;\om,\beta)$ defined by the following operational
transformation: \beq\label{defmomega1}
m_\La^\om=e^{-\frac{1}{4\om}\Delta}m_\La \qquad\mbox{and}\qquad
p_\La^\om=e^{-\frac{1}{4\om}\Delta}p_\La\,.\eeq We stress that
these superpolynomials are  two-parameter deformations of the
monomial and power-sum basis respectively. In that sense, a more
precise notation certainly would be $m_\La^{\om,\beta}$ and
$p_\La^{\om,\beta}$. Nevertheless, we stick to $m_\La^{\om}$ and
$p_\La^{\om}$ to lighten the notation and because the polynomials
(\ref{defmomega1}) satisfy, like the $\mc{D}_j^\om$'s and the
$J_\La^\om$'s,  the following limiting relation: \beq
\lim_{\om\rightarrow\infty}m_\La^\om=m_\La(x,\theta),\qquad\lim_{\om\rightarrow\infty}p_\La^\om=p_\La(x,\theta)\,.\eeq
Now, the transformation (\ref{eqsimtransfree2}) implies that
\beq\begin{array}{ll} \ds
\mathrm{H}_n^\om\,m_\La^\om=\big(\sum_j\La_j^n\big)\,m_\La^\om,&\ds
\mathrm{I}_{n-1}^\om\,m_\La^\om=\big(\sum_{j=1}^{\overline{\underline{\La}}}\La_j^{n-1}\big)\,m_\La^\om,\cr
\ds
\mathrm{H}_n^\om\,p_\La^\om=\big(\sum_j\La_j^n\big)\,p_\La^\om,&\ds
\mathrm{I}_{n-1}^\om\,p_\La^\om=\big(\sum_{j=1}^{\overline{\underline{\La}}}\La_j^{n-1}\big)\,p_\La^\om,
\end{array}\eeq
where $\La_i^n=(\La_i)^n$.  More generally, any symmetric and
homogenous superpolynomial $f(x,\theta)$ is related to a wave
function of the srCMS model {\it via} $\ds
e^{-\Delta/4\om}f(x,\theta)$. Since
$\{\mathrm{H}_n^\om,\mathrm{I}_{n-1}^\om\}$ is another complete
set of commuting operators in the srCMS model, the sets
$\{m_\La^\om\}_\La$ and $\{p_\La^\om\}_\La$ are new families  of
eigenfunctions.

\begin{theorem}\label{theodefmomega2}
The set $\{m_\Lambda^\om\}_\La$ is the unique basis of $P^{S_N}$
such that
  \beq \label{defm3}
\mc{H}_1^\om m_\Lambda^\om=|\La|m_\La^\om\qquad{and}\qquad
m_\La^\om=m_\La+\sum_{\Om<_u\La}y_{\La\Om}(\beta,\om,N)\,m_\Om.\eeq
Similarly, $\{p^\om_\La\}_\La$ is the only basis of $P^{S_ N}$
such that \beq \label{defp3} \mc{H}_1^\om
p_\Lambda^\om=|\La|p_\La^\om\qquad{and}\qquad
p_\La^\om=p_\La+\sum_{\Om<_u\La}z_{\La\Om}(\beta,\om,N)\,p_\Om.\eeq
\end{theorem}
\n{\it Proof.}  See Theorem
\ref{theodefhermite3}.\hfill$\square$\\

Notice that the functions $p_\La^\om$ generalize, in the
superspace, the solutions published in \cite{Brink} and
independently in \cite{UWa5}.  When $\om=1$ and the $\theta_i$'s
are replaced by the $\theta_i^\dagger$'s, the solutions $\ds
p^\om_\La=e^{-\Delta/4\om}p_\La$ are equivalent to the algebraic
constructions of Ghosh \cite{ghosh} (even though he did not use a
superpartition labelling).

We end this section by writing each $\om$-deformed superpolynomial
as a non-deformed superpolynomial, now expressed in terms  of $N$
differential operators rather than the $x_j$'s, acting on the
identity. In the zero-fermion sector, these algebraic expressions
reduce to those obtained in \cite{UWa4}.

\begin{proposition}Let $\hat{D}_i^\dagger=D_i^\dagger/2\om$ and
$\hat{D}^\dagger=(\hat{D}_1^\dagger,\ldots,\hat{D}_N^\dagger)$,
where $D_i^\dagger$ is the Dunkl operator defined in (\ref{Ddag}).
Then, \beq J^\om_\La=J_\La(\hat{D}^\dagger,\theta;1/\beta)\cdot
1\, ,\qquad m_\La^\om=m_\La(\hat{D}^\dagger,\theta)\cdot1\,
,\qquad\mbox{and}\qquad
p_\La^\om=p_\La(\hat{D}^\dagger,\theta)\cdot1\, .\eeq
\end{proposition}
\n{\it Proof.}  A direct calculation gives $[\Delta, x_j]=2D_i$,
hence (using again the
Baker-Campbell-Hausdorff formula)
\beq
e^{\frac{1}{4\om}\Delta}\,D_i^\dagger\, e^{-\frac{1}{4\om}\Delta}
=e^{\frac{1}{4\om}\Delta}\,(2\om x_i-D_i)\,
e^{-\frac{1}{4\om}\Delta}=2\om x_i.\eeq Thus, using
$\Delta\cdot 1=0$,  we get \beq m_\La(\hat{D}^\dagger,\theta)\cdot
1=m_\La\left(e^{-\frac{1}{4\om}\Delta}xe^{\frac{1}{4\om}\Delta},\theta\right)\cdot
1
=e^{-\frac{1}{4\om}\Delta}m_\La(x,\theta)e^{\frac{1}{4\om}\Delta}\cdot
1=m_\La^\om(x,\theta),\eeq
and similar relations hold for $J_\La(\hat{D}^\dagger,\theta;1/\beta)\cdot
1$ and $p_\La(\hat{D}^\dagger,\theta)\cdot 1$.\hfill$\square$\\

\section{Conclusion}

In this work, we have defined the generalized Hermite
superpolynomials as the eigenfunctions of the srCMS model that
decompose triangularly in terms of
  Jack superpolynomials.
The algebraic construction given in Section 4.1 amounts to
defining the generalized Hermite superpolynomials $J^\om_\La$ from
the Jack polynomials $J_\La$ as follows:
\begin{equation}
  J_\La^\om= e^{-\Delta/4\om} J_\La.
\end{equation}
This almost immediately implies their orthogonality. Albeit
elegant, a drawback of this construction should be pointed out:
the underlying computation depends on the number $N$ of variables
and does not directly express  the $J_\La^\om$'s in terms of the
Jack polynomials $J_\La$. Note also that the $J^\om_\La$'s are not
stable with respect to the number of variables.

In the same way as the Jack polynomials can be viewed as a
one-parameter deformation of the monomial basis that preserves
their orthogonality, the generalized Hermite superpolynomials are
thus obtained from  a step further orthogonality-preserving
deformation:
\begin{equation}\label{betaomegadef}
  m_\La \stackrel{\beta}{\longrightarrow} J_\La
\stackrel{\omega}{\longrightarrow} J_\La^\omega.
\end{equation}
Recall that the variables $x_j$ are the `position vectors': they
are constrained to the unit circle for the $m_\La$'s and the
$J_\La$'s while they lie on an infinite line for the
$J_\La^\om$'s.   The weight function also changes from one scalar
product to the other. Consequently, the scalar product is modified
at each step in Eq. (\ref{betaomegadef}).

As reviewed in Section 2.2, a symmetric polynomial in superspace
can be decomposed in terms of functions built out of a monomial in
fermionic variables times a polynomial of the bosonic variables
having mixed symmetry property. This indicates that a given type
of  superpolynomial could be reconstructed by an appropriate
`supersymmetrization' of the corresponding non-symmetric version
of the polynomial multiplied by suitable fermionic monomials. We
showed in \cite{DLM3} that the Jack superpolynomials can indeed be
constructed in this way. This route provides another method for
constructing the generalized Hermite superpolynomials out of the
non-symmetric generalized Hermite polynomials defined in
\cite{BF2}.

Still another construction is  presented in \cite{DLM5},
in the spirit of \cite{LapointeLascouxMorse} and \cite{DLM2}.
It furnishes a different construction that expresses the generalized
Hermite superpolynomials in determinantal form.

An immediate extension of this work is to study in detail the
continuous-spectrum limit  $\omega=0$: this transforms the
generalized Hermite superpolynomials into  generalized Bessel
functions in superspace. Another very natural generalization is to
construct the eigenfunctions of the supersymmetric CMS models
rational or trigonometric defined for other root systems. At first
sight the lift to other root systems appear to be  mostly a
technical problem.  One can however identify a more difficult problem
along these `extension lines'. It is the construction of the
supersymmetric version of the relativistic CMS models and their
eigenfunctions, the Macdonald superpolynomials.

Finally, let us stress that as an aside of our study, we have
clarified the integrability structure of the srCMS model by
expressing its conserved charges in terms of a similarity
transformation of the stCMS ones or the free ones. Moreover, we
have constructed supplementary conserved quantities whose
existence makes the model superintegrable.

\medskip
\medskip
\begin{acknow}
This work was  supported by NSERC and FONDECYT (Fondo Nacional de
Desarrollo Cient\'{\i}fico y Tecnol\'ogico) grant \#1030114. L.L.
wishes to thank Luc Vinet for his support at the early stages of
this work and P.D. is grateful to the Fondation J.A.-Vincent for a
student fellowship.
\end{acknow}

\end{document}